# QCD studies at LEP I

Giuseppe Raso

I.N.F.N. Bari

## Abstract

The high hadronic event statistics collected at the Z energy (LEP I) allowed a good understanding of the QCD dynamics. The coupling constant $\alpha_s$ has been measured with several methods giving a global average $\alpha_s(M_Z) = 0.122 \pm 0.004$. The flavour independence of $\alpha_s$ has been tested obtaining $\alpha_s^b/\alpha_s^{udsc} = 0.997 \pm 0.023$. Quark-gluon jet differencies has been observed among which $< n >_{gluon} / < n >_{quark} = 1.234 \pm 0.027$. A big role has been plaied by the silicon vertex detectors.

# 1  Introduction

The Quantum Chromo Dynamics (QCD)[1] is the most successful theory describing the strong interaction of quarks. Its perturbative version (PQCD) has been exploited to describe a large amount of data collected since decades. The only free parameter of the theory, the coupling constant $\alpha_s$, has been measured with accuracy limited so far only by theoretical uncertainties. Once the value of $\alpha_s$ is established, QCD can be tested comparing the predictions to the available experimental data. In particular, the large statistics available at LEP allows stringent tests of QCD. Tests on the gluon spin, gauge structure, running of $\alpha_s$, flavour independence of $\alpha_s$, differences between quark and gluon jets and other tests have been performed at LEP; in most cases the precision attaint before the LEP advent has been crucially improved.

In this talk I shall briefly report on the status of the $\alpha_s$ measurements at LEP (sect.2). Then I'll discuss in detail the experimental investigations on two important properties of QCD: the flavour independence of $\alpha_s$ (sect.3) and the quark-gluon jet differences (sect.4). In particular, I'll show that substantial improvements in the understanding of these aspects have been obtained thanks to the excellent features and performances of the LEP detectors and to the improved methods of analysis.

# 2  Status of $\alpha_s$ measurements at LEP

The advent of the e$^+$e$^-$ collider LEP working around the Z peak allowed a sizeable improvement in the tests of QCD and, in particular, in the measurement of $\alpha_s$. Actually, this measurement has been performed with different methods, at two energies ($M_Z$ and $M_\tau$) and for different quark flavours in the same experiment, allowing to test, respectively, the *consistency*, *running* and *flavour independence* of the coupling constant. Essentially two kinds of methods are employed at LEP to determine $\alpha_s(M_Z)$:

a) methods based on counting of the events

b) methods based on the analysis of the event topology.

About the method a) at LEP I it was possible to determine $\alpha_s$ from the ratio $R_{Z,\tau}$ of the hadronic to leptonic partial decay widths of Z and $\tau$ lepton:

$$R_{Z,\tau} = \frac{\Gamma_{had}}{\Gamma_{lep}} = R^0_{Z,\tau}(1 + \delta^{pert}_{Z,\tau} + \delta^{non-pert}_{Z,\tau}).$$

where $R^0_{Z,\tau}$ is the purely electroweak part, $\delta^{pert}_Z$ is the perturbative QCD correction and $\delta_Z^{non-pert}$ is the non-perturbative correction.

The measurements based on the method a) generally provided the most accurate determinations of $\alpha_s$. As a matter of fact $\delta_{Z,\tau}^{pert}$ are kwown to $O(\alpha_s^3)$, the non-perturbative effects are negligibles or smalls and the statistics collected is very high.

About $R_Z$, using the most recent results from the LEP experiments [3] and the theoretical prediction given in [2], one obtains:

$$\frac{\Gamma_{had}}{\Gamma_{lep}} = 20.788 \pm 0.032$$

from which one obtains

$$\alpha_s(M_Z) = 0.125 \pm 0.006$$

For the determination of $\alpha_s$ from $R_\tau$ [14] the non-perturbative part was estimated to be $\delta_{non-pert} = -0.007 \pm 0.004$, while $\delta_{pert}$ has been computed again to complete $O(\alpha_s^3)$.

Experimentally $R_\tau$ is obtained from the ratio :

$$R_\tau = \frac{1 - B_e - B_\mu}{B_e}$$

Averaging the two LEP results [15] one obtains:

$$\alpha_s(M_\tau) = 0.361 \pm 0.023$$

This measurement supports the $\alpha_s$ running predicted by QCD and again averaging the quoted translated values for $\alpha_s(M_Z)$ one obtains:

$$\alpha_s(M_Z) = 0.122 \pm 0.003$$

where the main contribution to the error comes from the theoretical uncertainties.

As far as the method b) is concerned, many *infrared* and *collinear safe* variables have been employed. These variables describe the event shape and are sensitive to the gluon radiation: Thrust, C-parameter, Differential 2-jets rate($D_2$), Energy-Energy correlation, Jet broadening mass, Oblateness and so on. The measurement of these variables is affected by hadronization corrections which cannot be computed perturbatively because they involve an energy scale around 1 GeV, where $\alpha_s$ is no longer small. Then, for the non-perturbative part one must rely on phenomenological approaches based on Monte-Carlo models such as Jetset[34], Herwig[35] and Ariadne[36]. The perturbative part has been computed at the $O(\alpha_s^2)$ and, more recently [4], using resummed NLLA + $O(\alpha_s^2)$ calculations. A summary of $\alpha_s(M_Z)$ from the analysis of the event shape variables is shown in table 1 and in fig.1. The main contribution to the total error comes from the hadronization correction and from the theoretical uncertainties.

| Experiment | $\alpha_s(M_Z)$ | Theory | Reference |
|---|---|---|---|
| ALEPH | $0.117^{+0.008}_{-0.010}$ | $O(\alpha_s^2)$ | [6] |
| DELPHI | $0.113 \pm 0.007$ | $O(\alpha_s^2)$ | [7] |
| L3 | $0.118 \pm 0.010$ | $O(\alpha_s^2)$ | [8] |
| OPAL | $0.122^{+0.006}_{-0.005}$ | $O(\alpha_s^2)$ | [9] |
| ALEPH | $0.125 \pm 0.005$ | $O(\alpha_s^2)$ + NLLA | [10] |
| DELPHI | $0.123 \pm 0.006$ | $O(\alpha_s^2)$ + NLLA | [11] |
| L3 | $0.124 \pm 0.009$ | $O(\alpha_s^2)$ + NLLA | [12] |
| OPAL | $0.120 \pm 0.006$ | $O(\alpha_s^2)$ + NLLA | [13] |

Table 1: $\alpha_s$ measurements at LEP from event shape variables.

A method has been proposed in ref.[5] to compute a global average of measurements from the 4 different experiments even though the exact correlation pattern is unknown. Applying this method to the results shown in table 1 one obtains :

$$\alpha_s(M_Z) = 0.121 \pm 0.005$$

The measurement of $\alpha_s$ based on method b) will be gold plated at the energy of W's (LEP II) where the measurements based on method a) will are affected by a large statistical error.

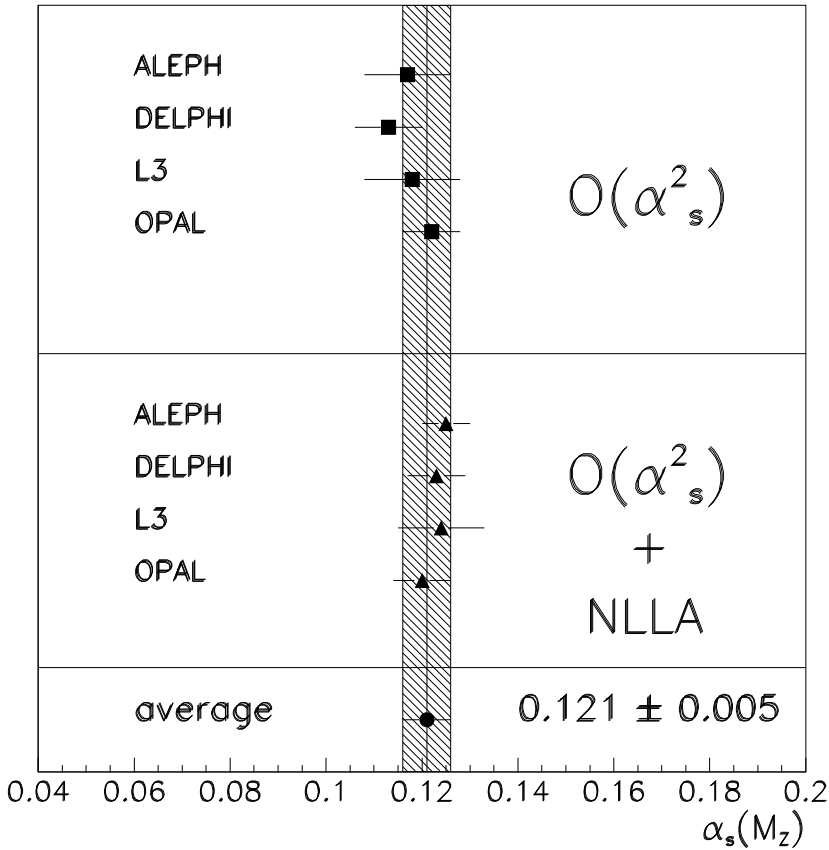

Figure 1: LEP measurement of $\alpha_s$ from event shape variables.

All these measurements of $\alpha_s(M_Z)$ from different techniques are in very good agreement, as one can see in table 2 where also the global average

$$\alpha_s(M_Z) = 0.122 \pm 0.004$$

computed according the method in ref.[5] is reported.

## 3 Test of flavour independence

In QCD the dynamics of the strong interaction is described by the lagrangian density

$$\mathcal{L} = \bar{q}_\alpha^{a,j}[i\gamma_{\alpha\beta}^\mu(\delta_{ab}\partial_\mu + igT_{ab}^r A_\mu^r) - M_j\delta_{ab}\delta_{\alpha\beta}]q_\beta^{b,j} - \frac{1}{4}F_{\mu\nu}^r F^{r,\mu\nu} \; .$$

where

$$F_{\mu\nu}^r = \partial_\mu A_\nu^r - \partial_\nu A_\mu^r - gf^{rst}A_\mu^s A_\nu^t.$$

| Method | $\alpha_s(M_Z)$ |
|---|---|
| $R_Z$ | $0.125 \pm 0.006$ |
| event shape variables | $0.121 \pm 0.005$ |
| $\tau$ hadr. decays | $0.122 \pm 0.003$ |
| Global average | $0.122 \pm 0.004$ |

Table 2: $\alpha_s$ measurements at LEP from different methods.

The same coupling constant $g$ appears in the quark-gluon and in three and four gluon vertices. So, unless to have gluons of different flavour, QCD predicts that $\alpha_s$ is independent of the quark flavour.

The agreement of the $\alpha_s$ values from various measurements done in the past in very different hadronic environment is already an indication of the flavour independence of $\alpha_s$, due to the different flavour composition involved. Moreover some dedicated measurements have been performed in order to test this particular property of QCD. In the past years test on the flavour independence of $\alpha_s$ have been performed studying the quarkonium states [16], the bottom production at p$\bar{\text{p}}$ colliders [17] and the relative strengths for charm and bottom quark measured in e$^+$e$^-$ colliders at centre-of-mass energies 30 GeV [18].

At LEP the particular conditions of the process e$^+$e$^-$ $\to$ Z $\to$ q$\bar{\text{q}}$, as well as the almost complete hermeticity of the detectors on the solid angle together with the last generation of silicon vertex detectors, allowed an almost complete reconstruction of the event. In such conditions a substantial improvement of this measurement has been obtained.

## 3.1 *B*-tagging procedures

At LEP energies the dominant process is the production of the vector boson Z which decays in q$\bar{\text{q}}$ pairs in all the available flavours almost democratically. The test of the strong interaction flavour independence can be exploited using only events originated by a specific flavour. That implied a tagging procedures that allow the separation of the different flavours. In particular, to separate the *b* quark from the other quarks two methods, based on different event signatures, have been used in the LEP experiments: a) lepton tagging and b) lifetime tagging.

### 3.1.1 Lepton tagging

This method exploits the feature of the semileptonic decays of heavy quarks of yielding prompt leptons with high momentum and high transverse momentum which can be used to identify b$\bar{\text{b}}$ events. In fact the hard fragmentation function of the *b* quark, to respect to lighter quarks, generally provides a *b* hadron with high momentum; moreover, the heavy *b* hadron mass gives leptons with high momentum to respect to the jet axis.

In ALEPH [19], for example, the procedures adopted to identify the *electrons* make use of the $dE/dx$ measurement in the TPC as well as the shape of the showers in the electromagnetic calorimeter. The *muons* are identified using the tracking capabilities of the hadron calorimeter together with information from the muon chambers.

By applying typical cuts of 4 GeV (3 GeV) on the momentum and of 1.5 GeV (1.0 GeV) on the transverse momentum of muons (electrons) in a hadron selected sample, the *b*-purities at LEP range between 60% and 80% with efficiencies of about 5-10%.

### 3.1.2 Lifetime tagging

The advent of high precision silicon vertex detectors has opened an alternative possibility for the *b*-tagging by looking at the experimental signature of the relatively long lifetime of the *b* hadrons. The high precision achieved on the impact parameter determination (about 25 $\mu$m for high momentum charged tracks) allowed to obtain a hadron sample with very high *b*-purities without penalizing the efficiencies, as shown in fig.2 where the purity/efficiency curve for *b*-tagging is plotted for the two tagging methods. In most case the discriminant variable used is the impact parameter significance $S$, defined as the signed impact parameter divided by its measurement error. The $S$ evaluation requires accurate estimates of the particle trajectory, Z decay vertex and errors on these quantities.

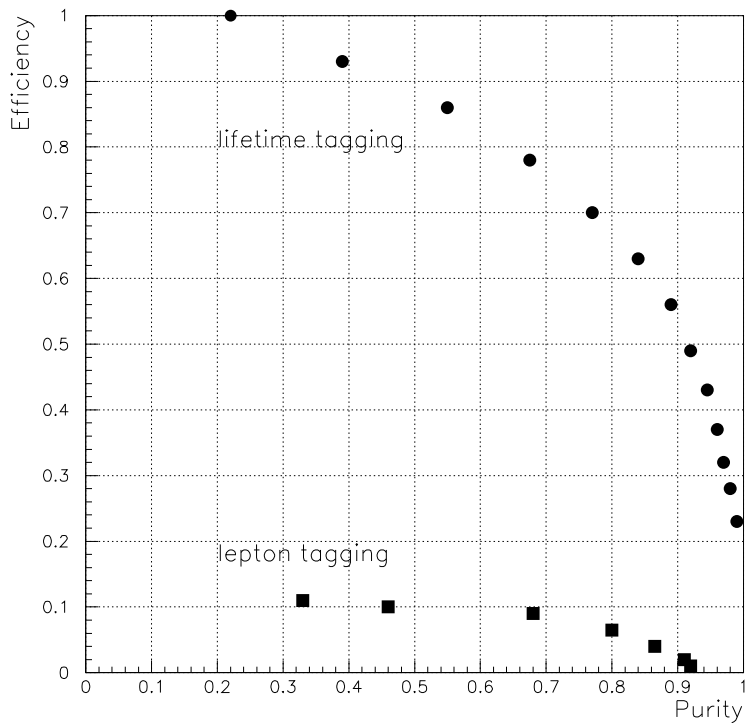

Figure 2: Purity vs. efficiency curve for $b$ tagging using the lifetime or the lepton method at ALEPH.

Typical $b$-purities obtained by this method at LEP are of about 90% for efficiencies of about 50% and light quark(u,d,s) contamination of about 0.5%.

## 3.2 Measurements

The first measurement was done by L3 [21] using a sample of b quark selected by the lepton tagging method. By applying cuts of 4 GeV (3 GeV) on the momenta and of 1.5 GeV (1.0 GeV) on the tranvserve momenta of muons (electrons) in a hadron sample of 110000 events L3 obtains a b-enriched sample with about 86% (88%) of purity.

Then the ratio $R_3$ of the 3-jet rates obtained for the two samples with the E0 jet finder algorithm (tab.3), have been evaluated:

$$\frac{R_3^{tag}}{R_3^{untag}} = \frac{R_3^b \beta + R_3^{udsc}(1-\beta)}{R_3^b \gamma + R_3^{udsc}(1-\gamma)}$$

where $\beta$ and $\gamma$ denote the b-purity in the b-enriched sample and in the complete hadronic sample respectively. Before translating this measurement in a measurement of $\alpha_s^b/\alpha_s^{udsc}$ some correction factors to the data are needed. L3 takes into account the corrections due to the hadronization (3% for muons, 7% for electrons), mass effects (2%) and detector acceptance and resolution (3%).

| Algorithm | Resolution | Recombination |
|---|---|---|
| Durham($k_T$) | $y_{ij} = \frac{2 \cdot min(E_i^2, E_j^2) \cdot (1 - cos\theta_{ij})}{E_{vis}^2}$ | $p_k = p_i + p_j$ |
| E0 | $y_{ij} = \frac{(p_i + p_j)^2}{E_{vis}^2}$ | $\vec{p_k} = \frac{E_k}{|\vec{p_i} + \vec{p_j}|}(\vec{p_i} + \vec{p_j})$<br>$E_k = E_i + E_j$ |
| E | $y_{ij} = \frac{(p_i + p_j)^2}{E_{vis}^2}$ | $p_k = p_i + p_j$ |
| P | $y_{ij} = \frac{(p_i + p_j)^2}{E_{vis}^2}$ | $\vec{p_k} = \vec{p_i} + \vec{p_j}$<br>$E_k = |\vec{p_k}|$ |
| Jade | $y_{ij} = \frac{2(E_i E_j)(1 - cos\theta_{ij})}{E_{vis}^2}$ | $p_k = p_i + p_j$ |
| Geneva(G) | $y_{ij} = \frac{8(E_i E_j)(1 - cos\theta_{ij})}{9(E_i + E_j)^2}$ | $p_k = p_i + p_j$ |

Table 3: Some jet finder algorithm definitions.

In fig.3 the ratio
$$\frac{R_3^b}{R_3^{udsc}} = \frac{\alpha_s^b}{\alpha_s^{udsc}}$$
is shown vs. $y_{cut}$ (the minimum jet resolution cut-off) after applying the correction factors. It should be noticed that this relation is only true at the first order in $\alpha_s$; in the L3 analysis the second order corrections are considered to be negligible. By taking the value of this ratio at $y_{cut}$=0.05 L3 obtains:
$$\frac{\alpha_s^b}{\alpha_s^{udsc}} = 1.00 \pm 0.05(stat) \pm 0.06(syst)$$
where the systematical error is due to MonteCarlo statistics(0.05), detector correction(0.03)

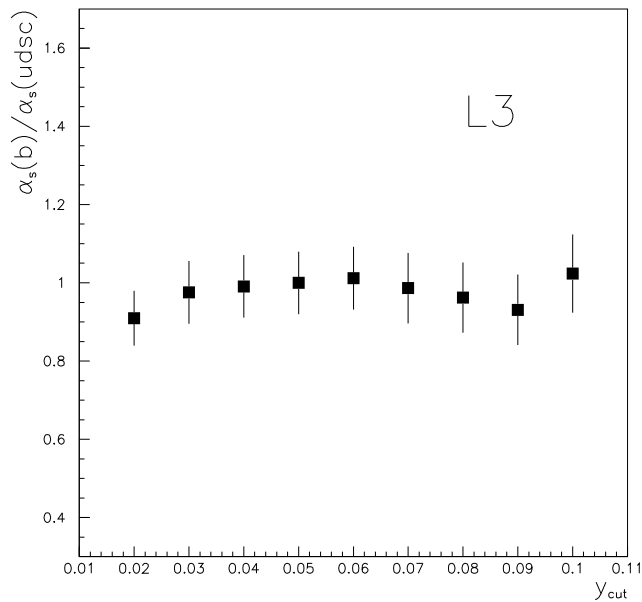

Figure 3: The ratio $R_3^b/R_3^{udsc}$ vs. $y_{cut}$ in L3 analysis.

and hadronization correction(0.02).

A similar analysis was performed by DELPHI[22] collaboration using a sample of 356000 hadronic events. They obtain a b-enriched sample, using the lepton tagging method, with a purity of 76%(68%) for muon(electron) sample by cutting at 4 GeV (3 GeV) on momentum and 1.5 GeV (1.5 GeV) on transverse momentum of muons(electrons). As in the L3 analysis, the variable used is the 3-jet rate but with four different jet finder algorithms: E0,P,$k_T$,G (tab.3). They apply the corrections both to the data (detector resolution,hadronization) and to the theoretical predictions (cuts bias, mass effects). Particular attention was paid to the cut bias correction factor, by computing different coefficients for each channel producing leptons. About the mass effect corrections they use two possible choices: an $O(\alpha_s)$ correction and a weigthed second order correction $W_1 O(\alpha_s) + W_2 O(\alpha_s^2)$.

In fig.4 the ratio $R_3(b)/R_3(udsc)$ vs. $y_{cut}$ is reported for each metric scheme; the effect of the mass corrections is also shown. The result at $y_{cut} = 0.06$ is:

$$\frac{\alpha_s^b}{\alpha_s^{udsc}} = 0.97 \pm 0.04(stat) \pm 0.04(syst)$$

where the main systematic contribution comes from MonteCarlo statistics and from cut bias extimation. In the same paper an alternative method is presented using a likelihood fit from $p$ and $p_T$ distributions of the leptons in 2 and 3 jets events, the result in this case

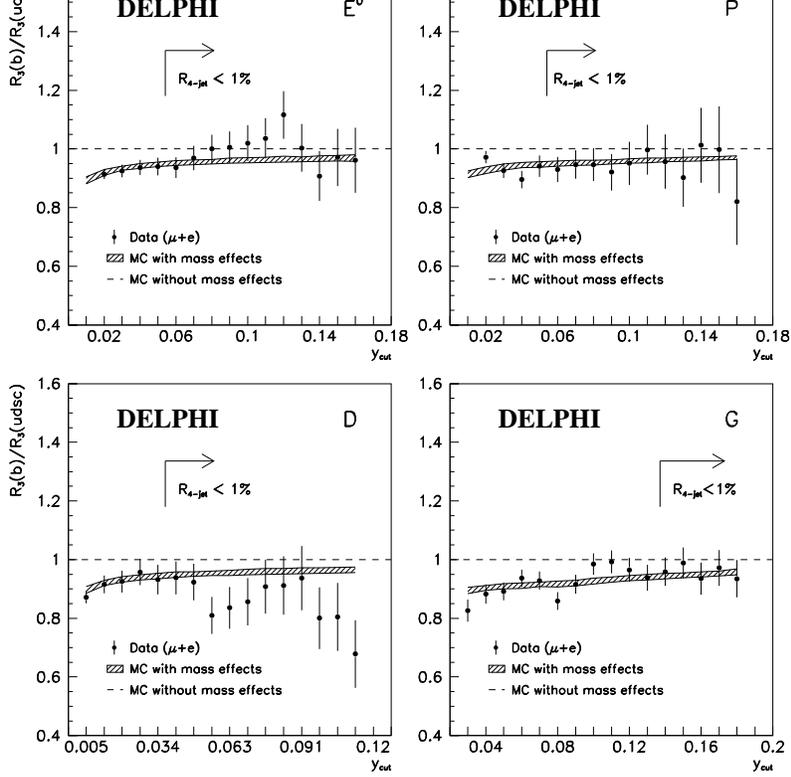

Figure 4: The ratio $R_3^b/R_3^{udsc}$ vs. $y_{cut}$ for 4 different jet finder algorithms by DELPHI.

is:

$$\frac{\alpha_s^b}{\alpha_s^{udsc}} = 1.00 \pm 0.04(stat) \pm 0.03(syst).$$

OPAL published two papers on this item; in the first one [23] a complete analysis is done by selecting different samples enriched in $b$, $c$, $s$ or $uds$ flavours. This is obtained from a sample of about 630000 hadronic events tagging the different flavours by requiring high momentum and high transverse momentum particles: leptons, $D^*$ or $K_S^0$ for $b$, $c$ and $s$ enriched sample respectively. On the other hand, the $uds$ enriched sample is obtained selecting events with high $x = 2E/E_{cm}$ tracks. The purities for each sample are given in table 4.

The variable used is the differential 2-jet rate distribution

$$D_2(y) = \frac{R_2(y) - R_2(y - \Delta y)}{\Delta y}$$

$D_2$ is the distribution of the jet resolution parameter $y$ at which 3-jet events turn into 2-jet events. Using this variable instead $R_3$ the bin correlation decreases because each event contribuites only once to the distribution. OPAL chooses to apply almost all the corrections to the data distributions; using this procedure one has to take into account also the flavour composition of the data samples obtained by MonteCarlo models. So the correct $D_2$ distribution for the flavour $f$ is

$$D_{2,cor}^{f}(y_i) = \Sigma_j C^f(y_i, y_j)[D_2^{f,obs}(y_j) - D_{2,MC}^{compl}(y_j)]$$

where the correction matrix $C^f(y_i, y_j)$ is evaluated by MonteCarlo, taking into account :

- biases due to the tagging procedure
- distortion due to the limited acceptance and resolution of the detector
- hadronization effects
- initial state radiation

| Flavour | $\mu$ sample | $e$ sample | $D^*$ sample | $K_S^0$ sample | High $x$ sample |
|---|---|---|---|---|---|
| u | 2.2 ± 1.0 | 1.3 ± 1.0 | 4.5 ± 4.0 | 8.7 ± 3.5 | 30.1 ± 4.1 |
| d | 2.2 ± 1.0 | 1.3 ± 1.0 | 4.5 ± 4.0 | 15.8 ± 3.5 | 28.7 ± 3.9 |
| s | 2.2 ± 1.0 | 1.3 ± 1.0 | 4.5 ± 4.0 | 53.6 ± 3.5 | 30.6 ± 5.2 |
| c | 7.6 ± 1.7 | 9.6 ± 1.5 | 59.1 ± 5.6 | 16.0 ± 2.9 | 3.7 ± 1.4 |
| b | 85.8 ± 1.3 | 1.3 ± 86.5 | 27.4 ± 4.1 | 5.9 ± 2.1 | 6.9 ± 4.0 |

Table 4: Flavour purities (in %) of the tagged samples in OPAL.

The correction to $\alpha_s^b$ due to the quark mass effects is applied instead to the thoretical distributions using the calculation given in ref.[24]. Moreover in [25] OPAL measures $\alpha_s^b/\alpha_s^{udsc}$ from others event shape variables as jet masses, thrust and energy- energy correlation, using also the lifetime tagging. The results from OPAL are given in table 5 where the main systematic uncertainties come from MonteCarlo statistics, from tagging procedure and from the renormalization scale in the fit.

ALEPH has measured [26] the ratio $\alpha_s^b/\alpha_s^{udsc}$ comparing the event shape variables Thrust, C-parameter, $D_2^{Jade}$ and $D_2^{Durham}$ for a full hadronic sample (900000 events) and for a $b$-enriched sample. In order to minimize the systematic uncertainty two enriched samples have been obtained using the two tagging procedures described in 3.1.1 and 3.1.2, with purity of 88% and 86% respectively. As in the OPAL analysis, ALEPH uses the second order QCD prediction for the distribution of the variable $X$:

$$\frac{1}{\sigma_0}\frac{d\sigma}{dX} = \frac{\alpha_s(\mu^2)}{2\pi}A(X) + [\frac{\alpha_s(\mu^2)}{2\pi}]^2[A(X)2\pi b_0 ln\frac{\mu^2}{M_z^2} + B(X)]$$

where $A(X)$ and $B(X)$ are tabulated in [27] and $\mu$ is the renormalization scale set to $\mu^2 = 0.05 \cdot M_Z^2$ in the fit. The corrections taken into account are the same as before but now the correction factors are applied to the theoretical predictions so that the measured ratio

$$R_{data} = \frac{\frac{1}{N}\frac{dN}{dX}|_{tag}}{\frac{1}{N}\frac{dN}{dX}|_{Q\bar{Q}}}$$

is fitted to the theoretical expression

$$R_{th} = \frac{G_{tag}^b \cdot f_{tag}^b + G_{tag}^{udsc} \cdot (1 - f_{tag}^b)}{G_{Q\bar{Q}}^b \cdot f_{Q\bar{Q}}^b + G_{Q\bar{Q}}^{udsc} \cdot (1 - f_{Q\bar{Q}}^b)}$$

where $tag$ and $Q\bar{Q}$ denote the tagged and the full hadronic sample respectively, $f$ are the b-purity and $G$ are the theoretical functions unfolded by the full set of correction factors. Figure 5 shows the measured ratio $R_{data}$ for each variable, when the lepton tagging is used, compared to the fitted theoretical predictions $R_{th}$. By combining the results from each variable and from different tagging procedures and taking into account the correlation ALEPH obtains:

$$\frac{\alpha_s^b}{\alpha_s^{udsc}} = 1.002 \pm 0.009(stat.) \pm 0.005(syst.) \pm 0.021(theo.).$$

| Flavour | $\alpha_s^f/\alpha_s^{compl}$ | Tagging |
|---------|-------------------------------|---------|
| b | $1.017 \pm 0.036$ | leptons |
| b | $0.992\ ^{+0.015}_{-0.016}$ | lifetime |
| c | $0.918 \pm 0.115$ | $D^*$ |
| s | $1.158 \pm 0.164$ | $K_S^0$ |
| u,d,s | $1.038 \pm 0.221$ | High $x$ |

Table 5: The ratio of $\alpha_s$ values for different quark flavour in OPAL.

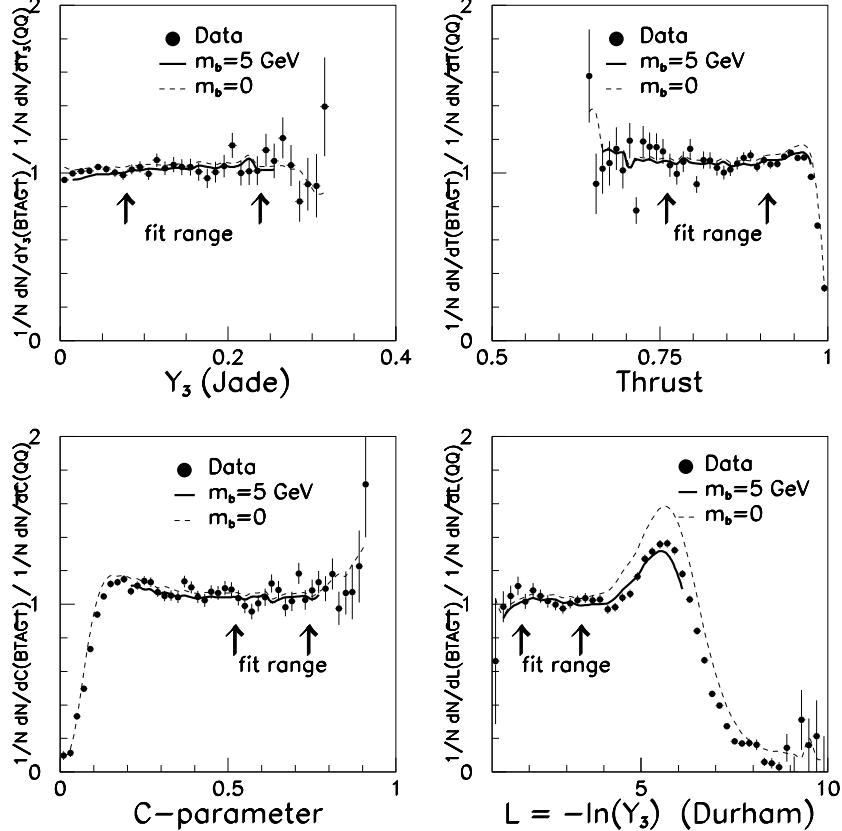

Figure 5: Ratio of the normalized cross section in ALEPH of the *b*-enriched sample tagged with high-$p_T$ lepton and the full hadronic sample. the full circles are the data, the solid line represents the fit result and the dashed line represents the theoretical prediction without the mass corrections.

Furthermore the lifetime tagging allows to select an *uds*-enriched sample; therefore ALEPH gives also the measurement of the ratio

$$\frac{\alpha_s^{uds}}{\alpha_s^{cb}} = 0.971 \pm 0.009(stat.) \pm 0.011(syst.) \pm 0.018(theo.).$$

The main systematic uncertainties come from mass correction, hadronization and renormalization scale; moreover, for the lifetime tagging, also the cut bias becomes important as it is explained later on.

To combine the previous results from L3, DELPHI, OPAL and ALEPH summarized in figs.6a and 6b we use the method given in [5] obtaining

$$\frac{\alpha_s^b}{\alpha_s^{udsc}} = 0.997 \pm 0.023$$

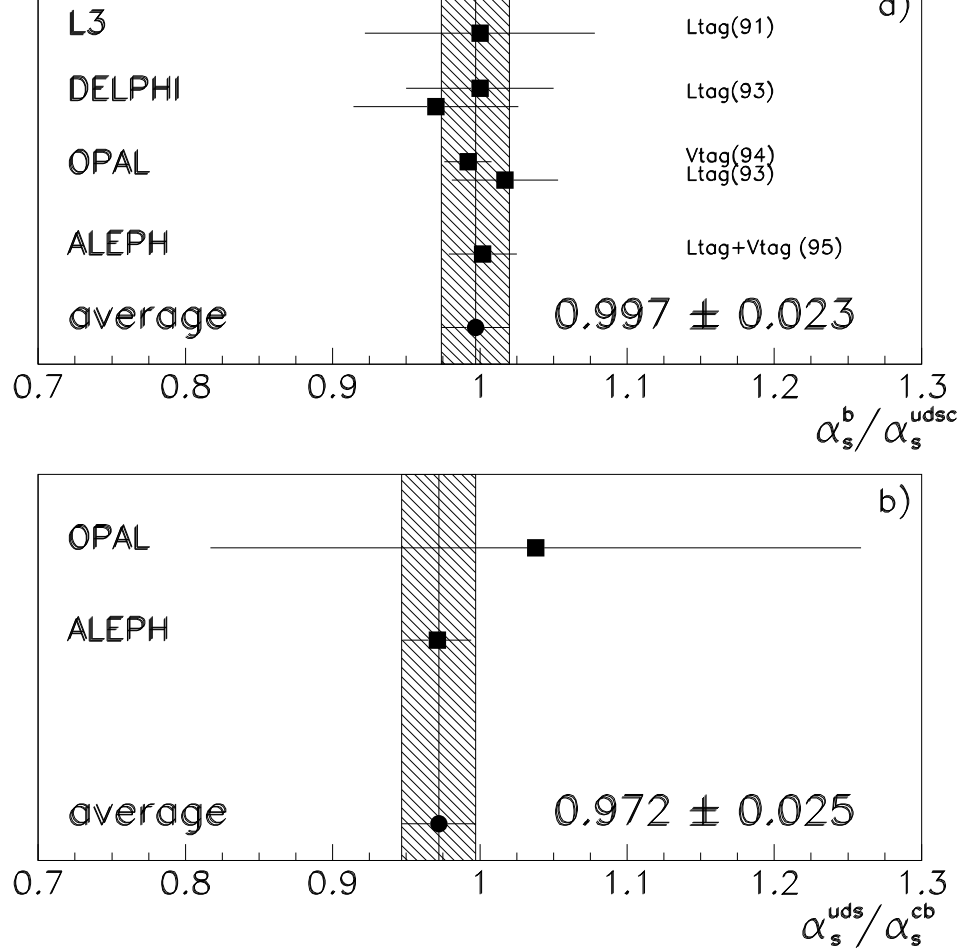

Figure 6: Compilation of the measurements of a) $\alpha_s^b/\alpha_s^{udsc}$ and b) $\alpha_s^{uds}/\alpha_s^{cb}$ at LEP.

$$\frac{\alpha_s^{uds}}{\alpha_s^{cb}} = 0.972 \pm 0.025$$

and from OPAL alone

$$\frac{\alpha_s^c}{\alpha_s^{udsb}} = 0.918 \pm 0.115.$$

These measurements represent the best world test of the flavour independence of the strong interactions.

## 3.3 Systematic uncertainties

The systematic errors can be divided in two categories : experimental and theoretical uncertainties. The procedures to determine the experimental uncertainties are the same for all experiment: normally they are evaluated by varying the parameters and the cuts used for data selection and tagging. Thanks to the good detector resolution, usually this part of systematic error contributes for less than 1%; the only exception is the bias due to the lifetime tagging because in this case the correction factors are not negligible. Notice that this feature is expected due to a greater 2-jet-like nature of the events with longer lifetime which are easier tagged.

Concerning the evaluation of the theoretical systematical errors, different procedures, more or less conservative, have been followed from each experiment. The effect of the hadronization, for instance, is important when one uses the event shape variables and moreover each MonteCarlo model gives different agreements for different variables. This imposed to use in the fitting procedures limited ranges where the hadronization corrections are minimal, and to evaluate the related uncertainty using as many MonteCarlo models as possible.

Another source of theoretical systematic error is related to the renormalization scale $\mu$ used in the fit. As it is well kwown, using the exact QCD second order predictions a small $\mu$ variation in the fit causes a notable $\alpha_s$ variation because this parameter takes partially into account the missing higher order contributions. However in the flavour independence test one deals with ratios of $\alpha_s$ and this presumably reduces the $\mu$ dependence; nevertheless the residual effect is not yet negligible. As stressed in [28], in evaluating this error one cannot choose a standard range of variation for the $\mu$ parameter, and each experiment makes different choices. We think that for the $\alpha_s^b/\alpha_s^{udsc}$ measurement the more conservative range for the $\mu$ parameter is between the $b$ quark mass to the Z mass. A way to reduce the systematical error should be to use the resummed NLLA + $O(\alpha_s^2)$ calculation in the theoretical predictions because this is known to reduce the $\mu$ depencence.

Furthermore, only the tree level second order mass corrections have been computed so far [24], and their use in the correction procedure is not obvious, so in certain cases [26] these calculations have been used for a rough estimation of the relative uncertainty. Using the complete second order mass corrections should reduce significantly the systematical error.

# 4 Properties of quark and gluon jets

According to QCD the quarks have a single color charge while the gluons carry two color indices; that causes a different coupling strength for quarks and gluons to emit an additional gluon as it is denoted by the Casimir factors $C_A$ and $C_F$: $C_A$=3 gives the relative strength for the gluon-gluon coupling while $C_F = \frac{4}{3}$ gives the quark-gluon coupling strength. As a conseguence one expects that the jets initiated by quark and gluons have different features, which could be experimentally observed. For instance the mean particle multiplicity ratio for gluon and quark jets is asymtotically expected to be

$$\frac{<n>_{gluon}}{<n>_{quark}} = \frac{C_A}{C_F} = 2.25$$

and at the next to leading order[31]

$$\frac{<n>_{gluon}}{<n>_{quark}} = 2.25[1 - 0.273 \cdot \sqrt{\alpha_s(Q)} - 0.071 \cdot \alpha_s(Q)].$$

At the LEP energies this simple prediction is expected to be significantly altered by coherence effects [29], which strongly suppress the fragmentation of the gluon in the 3-jet like events, and by the hadronization which in some cases can mask the perturbative quark-gluon difference; in any case, the ratio is predicted to significantly differ from the unity.

## 4.1 Gluon tagging

Several analyses have been made [30] to look for evidence of such jet differences but often the strong bias introduced in tagging precedures and, as pointed out in [31], the not properly inclusive analysis techiques have yield experimental results not easily comparable to the theoretical predictions.

First of all, it is necessary to define the jets and to assign each particle to a jet. Essentially two jet finder algorithms are used : the DURHAM (or $k_T$) and the JADE algorithms. They differ in the definition of the recombination scheme and of the jet resolution parameter as summarized in table 3.

Then for a comparison of quark and gluon jet properties one needs samples of quark and gluon jets of similar energies. For this reason symmetric jet event typologies as thats shown in fig.7 are selected: the "Mercedes type" events, with $\theta_1 \simeq \theta_2 \simeq \theta_3 \simeq 120^0$, or the "Y type" events, with $\theta_2 \simeq \theta_3 \simeq 150^0$.

At LEP experiments the use of the vertex detectors supplied a powerful tool to identify the quark jets with respect to the gluon jets. In fact one of the two lower energy jets (Mercedes) or two of the three jets (Y) can be tagged as heavy quark jet by requiring a displaced secondary vertex as already seen in 3.1.2. In this way the gluon jet is actually anti-tagged, obtaining virtually unbiased jet properties. In such way one obtains two sample of 3-jet events: a natural mixed sample without any tagging where the gluon purity is around 50% and a anti-tagged sample whith a high gluon purity. Of course the

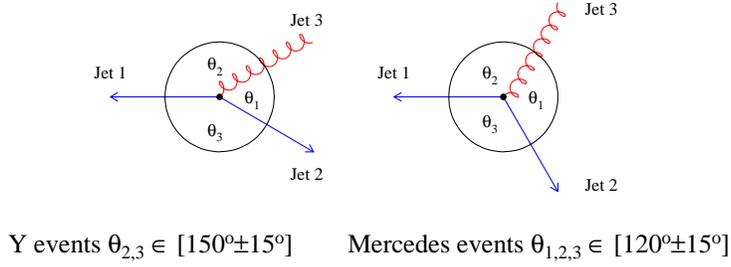

Y events $\theta_{2,3} \in [150°\pm15°]$     Mercedes events $\theta_{1,2,3} \in [120°\pm15°]$

Figure 7: Symmetric three-jet events of Mercedes or Y type.

flavour composition of the quark jet tagged sample is different from the natural mixing but the assumption is made that the anti-tagged gluon jet properties are independent of the quark flavour radiating it, as predicted by QCD [32].

Typical values of gluon purity and efficiency obtained at LEP using this method range between 70-90% and 4-10% respectively.

## 4.2 Measurements

In this type of analysis the comparison between a quark and gluon jet property $A_{q(g)}$ is normally unfolded by the correspondent property $A^{T(M)}$ of the tagged sample(T) and the natural mixture sample(M):

$$A^T = P_g^T A_g + (1 - P_g^T) A_q$$

$$A^M = P_g^M A_g + (1 - P_g^M) A_q$$

where $P_g^{T(M)}$ is the gluon purity in the T(M) sample.

Ideally the two samples should consist of events where the jets are produced in the same kinematical configurations and, as seen before, the tagging procedure should not introduce a bias. Correction procedures similar to the one adopted in 3.2 take into account the small (about 2%) biases introduced together with the detectors acceptance and resolution.

In ref.[33] OPAL reports the results for some quark and gluon jet properties by comparing the data to the predictions of Jetset[34], Herwig[35], Ariadne[36] and Cojets[37] parton shower models after tuning the parameters to provide a good description for the global event characteristics.

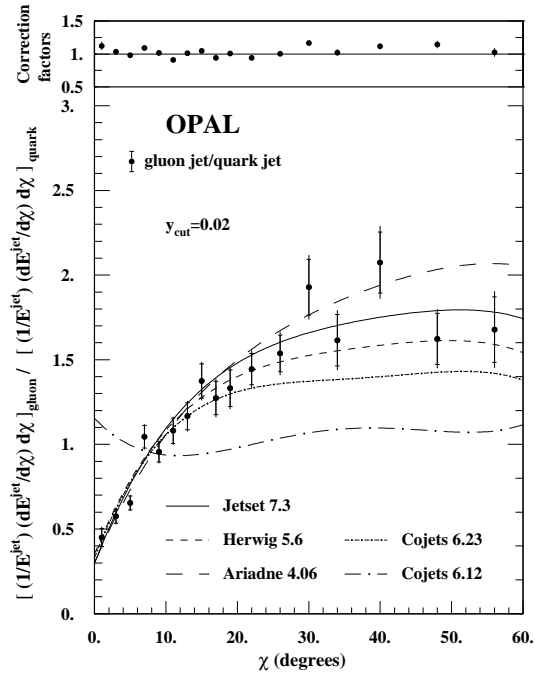

Figure 8: OPAL: the ratio of the distribution of the jet energy $E^{jet}$ with respect to the jet axis for gluon and quark jets versus the angle $\chi$ of a particle with respect to the jet axis.

One of the differences expected between quark and gluon jet is the angular distribution of the jet energy $E^{jet}$ with respect to the jet axis. Fig.8 shows the ratio of the gluon to the quark jet distributions $(1/E^{jet})(dE^{jet}/d\chi)d\chi$ versus $\chi$, where $\chi$ is the angle between a particle and the relative jet axis. The predictions of the various models are also shown and the Cojets 6.12 model is to be understood as a "toy model" since, in this version, no differences between quark and gluon jet are provided. Another feature expected to be differ in quark and gluon jets is the inclusive distribution of the particle energy in the jets, known as the fragmentation function. Fig.9 shows the ratio of the gluon to the quark jet distributions of the charged particle fragmentation function $(1/N_{event})dn_{ch}/dx_E$ versus $x_E = E/E^{jet}$. From figs.9 and 10 it is seen that the gluon jets are observed to be broader and to contain fewer energetic particles than quark jets as predicted by QCD; moreover the "toy model" is in evident disagreement with the experimental observations. Another important measurement performed by OPAL is the ratio of the mean particle multiplicity

$$\frac{<n_{ch}>_{gluon}}{<n_{ch}>_{quark}} = 1.25 \pm 0.02(stat.) \pm 0.03(syst.)$$

where the main contribution to the systematical error comes from experimental uncertainties and MonteCarlo statistics.

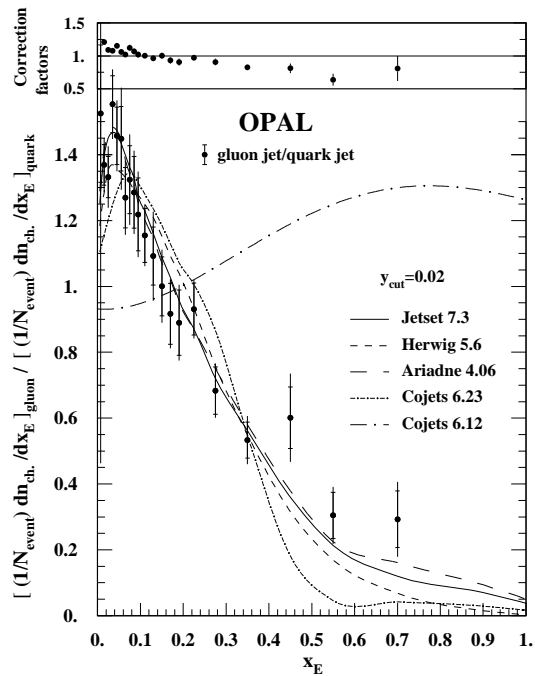

Figure 9: OPAL: the ratio of the charged particle fragmentation functions of gluon and quark jets.

ALEPH [38] gives a measurement of this quantity with a similar tagging and correction procedure as in OPAL. The value found is

$$\frac{<n_{ch}>_{gluon}}{<n_{ch}>_{quark}} = 1.19 \pm 0.04(stat.) \pm 0.02(syst.)$$

where the main systematical contributions are due to the tagging bias and to the detector corrections. ALEPH also gives [38] a measure of the ratio of the mean particle multiplicity of the gluon jet to the $b$ quark jet which has been found to be consistent with unity

$$\frac{<n_{ch}>_{gluon}}{<n_{ch}>_{bquark}} = 1.00 \pm 0.05(stat.) \pm 0.02(syst.).$$

This indicates that, for the energy scale involved, the additional particle multiplicity arising from the $b$ hadron decay masks completely the difference between $b$ quark and gluon jet multiplicity. This is in agreement with the OPAL result given in [39].

Another interesting way to look at the quark-gluon jet differences is given in [40] by ALEPH by studying the subjet structure of the jets. The method consist to analyse the subjet multiplicity of the quark ($N_q$) and gluon ($N_g$) jets by varing the resolution parameter $y_0$ of the jet finder algorithm after having selected three jet symmetric events by using the same algorithm with $y_1 > y_0$. In fig.10 the measured ratio $< N_g - 1 > / < N_q - 1 >$ is plotted versus $y_0$ together with the predictions of various MonteCarlo models. The behaviour is the result of both perturbative and non-perturbative effects where the last one becomes more important for small values of the resolution parameter.

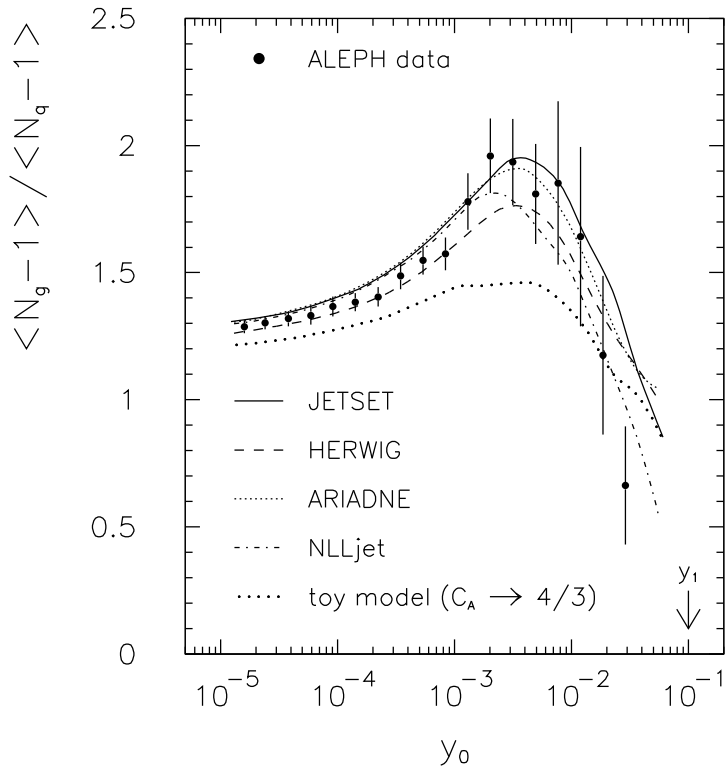

Figure 10: ALEPH: ratio of subjet multiplicities for gluon and quark jet.

An increase of the ratio $r = < n_{ch} >_{gluon} / < n_{ch} >_{bquark}$ with the energy has been reported by DELPHI [20]. The slope is found to be

$$\frac{\Delta r}{\Delta E} = (86 \pm 29(stat.) \pm 14(syst.)) \cdot 10^{-4} GeV^{-1}$$

to be compared with the Jetset hadron level value

$$\frac{\Delta r}{\Delta E} = (90 \pm 3) \cdot 10^{-4} GeV^{-1}$$

The indication for the energy dependence comes mainly from the comparison of non-symmetric $q\bar{q}g$ and $q\bar{q}\gamma$ events but is supported by the analysis of the symmetric events. From the study of the simmetric 3-jet events DELPHI obtains:

$$\frac{<n_{ch}>_{gluon}}{<n_{ch}>_{quark}} = 1.241 \pm 0.015(stat.) \pm 0.025(syst.)$$

A summary of the ratio for gluon to quark jet of the mean charged particle multiplicity is given in table 6 where the global average is also reported.

| Experiment | $\frac{<n_{ch}>_{gluon}}{<n_{ch}>_{quark}}$ |
|---|---|
| OPAL | $1.25 \pm 0.02 \pm 0.03$ |
| ALEPH | $1.19 \pm 0.04 \pm 0.02$ |
| DELPHI | $1.241 \pm 0.015 \pm 0.025$ |
| Global average | $1.234 \pm 0.027$ |

Table 6: The ratio for gluon to quark jet of the mean charged particle multiplicity.

All these measurements are in agreement with the QCD expectations; moreover, with respect to the quark jet, the gluon jet is seen to have higher particle multeplicity, softer fragmentation function and to be less collimated. Furthemore some analyses give evidence of a non-perturbative contribution to the quark-gluon difference.

# 5 Conclusions

The high statistics of the hadronic events from the Z decay collected by the LEP experiments allowed a remarkable understanding of the dynamics of QCD. Perturbative and non-perturbative aspects have been tested with good accuracy taking advantage from the high performance of the detectors.

The coupling constant $\alpha_s$ was measured using several independent methods obtaining a global average

$$\alpha_s(M_Z) = 0.122 \pm 0.004$$

by keeping under control the systematic uncertainties.

A test of the flavour independence of $\alpha_s$ was carried out thanks to the almost democratical Z decay and to several heavy flavour tagging techniques such as the lepton tagging and the lifetime tagging for the $b$ quark, giving as result

$$\frac{\alpha_s^b}{\alpha_s^{udsc}} = 0.997 \pm 0.023.$$

The high resolution of the silicon vertex detectors has supplied a powerful tool to separate gluon jet from quark jet, allowing a qualitative comparison of the data with several parton shower models. Also in this type of analysis the QCD predictions have been confirmed by observing a softer fragmentation function and a larger angular width of the gluon jet with respect to the quark jet. Moreover the ratio for gluon to quark jet of the mean charged particle multiplicity has been measured to be

$$\frac{<n_{ch}>_{gluon}}{<n_{ch}>_{quark}} = 1.234 \pm 0.027.$$

# 6 Acknowledgements

I would like to thank the colleagues Mauro de Palma and Pietro Colangelo for their generous and precious support in preparing this report.